\documentclass[sigconf]{acmart}

\setcopyright{none}                

\usepackage{subcaption}

\usepackage{amsmath,amsfonts}
\usepackage[algo2e,linesnumbered,ruled,lined,noend]{algorithm2e}
\SetKwInOut{Parameter}{Parameters}
\SetKwInOut{Constraint}{Constraints}
\SetKwFunction{Mining}{Mining}
\SetKwFunction{MiningTwo}{Mining2}
\SetKwFunction{Candidates}{Candidates}
\SetKwProg{Fn}{Function}{}{}
\usepackage{graphicx}
\usepackage{textcomp}
\usepackage{color}
\usepackage{xcolor}
\usepackage{dsfont}
\usepackage{bbm}

\usepackage[dvipsnames]{xcolor}

\usepackage{xspace}
\def\BibTeX{{\rm B\kern-.05em{\sc i\kern-.025em b}\kern-.08em
    T\kern-.1667em\lower.7ex\hbox{E}\kern-.125emX}}
    
\usepackage{multirow} 
\usepackage{booktabs} 
\usepackage{colortbl}

\usepackage[normalem]{ulem}
\useunder{\uline}{\ul}{}

\usepackage{pgf}
\usepackage{enumitem}
\setlist[itemize]{leftmargin=*}
\usepackage{stfloats}

\usepackage{makecell}

\usepackage{bbold}
\usepackage{mathtools}
\usepackage{comment}
\usepackage{microtype}
\usepackage[misc]{ifsym}
\usepackage{url}
\usepackage{xspace}

\SetCommentSty{mycommfont}

\newcommand{\smallsection}[1]{{\vspace{0.02in} \noindent {{\underline{\smash{\bf #1}}}}}}
\newcommand{\method}{{CoRRe}\xspace}

\definecolor{sunwoogreen}{RGB}{32, 200, 150}
\definecolor{sunwoogreen2}{RGB}{67, 148, 58}
\definecolor{sunwooyellow}{rgb}{1.0, 1.0, 0.0}
\definecolor{kyunghoblue}{RGB}{120, 170, 255}

\newcommand{\gframe}[1]{\fcolorbox{white}{sunwoogreen!70}{\strut #1}}
\newcommand{\yframe}[1]{\fcolorbox{white}{sunwooyellow!70}{\strut #1}}
\newcommand{\bframe}[1]{\fcolorbox{white}{kyunghoblue!40}{\strut #1}}
\usepackage{xcolor}
\usepackage{pifont}

\newcommand{\yes}{\textcolor{green!60!black}{\ding{51}}} 
\newcommand{\noo}{\textcolor{red}{\ding{55}}} 

\newcommand{\postllm}{\textcolor{orange}{\Large\ding{72}}} 
\newcommand{\prellm}{\textcolor{blue}{\ding{115}}} 

\AtBeginDocument{%
  \providecommand\BibTeX{{%
    \normalfont B\kern-0.5em{\scshape i\kern-0.25em b}\kern-0.8em\TeX}}}

\usepackage{needspace}



\copyrightyear{2026}
\acmYear{2026}
\setcopyright{cc}
\setcctype{by}
\acmConference[CIKM '26]{Proceedings of the 35th ACM International Conference on Information and Knowledge Management}{November 07--11, 2026}{Rome, Italy}
\acmBooktitle{Proceedings of the 35th ACM International Conference on Information and Knowledge Management (CIKM '26), November 07--11, 2026, Rome, Italy}
\acmDOI{10.1145/3799682.3839925}
\acmISBN{979-8-4007-2539-5/2026/11}

\begin{document}

   \title[Training-Free LLM-Based Recommendation with Post-LLM Item Refinement Using Collaborative Signals]{Training-Free LLM-Based Recommendation with \\ Post-LLM Item Refinement Using Collaborative Signals}

    \settopmatter{authorsperrow=3}

    \author{Kyungho Kim}	
    \affiliation{%
		\institution{KAIST}
            \city{Seoul}
            \country{Republic of Korea}
	}
	\email{kkyungho@kaist.ac.kr}
	
    \author{Sunwoo Kim}
	\affiliation{
    	\institution{KAIST}
            \city{Seoul}
            \country{Republic of Korea}
	}
	\email{kswoo97@kaist.ac.kr}

    \author{Geon Lee}
	\affiliation{
		\institution{KAIST}
            \city{Seoul}
            \country{Republic of Korea}%
	}
	\email{geonlee0325@kaist.ac.kr} 

    \author{Shinhwan Kang}
	\affiliation{
		\institution{KAIST}
            \city{Seoul}
            \country{Republic of Korea}%
	}
	\email{shinhwan.kang@kaist.ac.kr}

    \author{Sojeong Kim}
	\affiliation{
		\institution{KAIST}
            \city{Seoul}
            \country{Republic of Korea}%
	}
	\email{kimsojeong@kaist.ac.kr}

    \author{Liam Collins}
	\affiliation{
		\institution{Snap Inc.}
            \city{Bellevue, WA}
            \country{USA}%
	}
	\email{lcollins2@snapchat.com} 
    
    \author{Bhuvesh Kumar}
	\affiliation{
		\institution{Snap Inc.}
            \city{Bellevue, WA}
            \country{USA}%
	}
	\email{bkumar4@snapchat.com}

    \author{Donald Loveland}
	\affiliation{
		\institution{Snap Inc.}
            \city{Bellevue, WA}
            \country{USA}%
	}
	\email{dloveland@snapchat.com}
    
	\author{Kijung Shin}
	\affiliation{
		\institution{KAIST}
            \city{Seoul}
            \country{Republic of Korea}
	}
	\email{kijungs@kaist.ac.kr}
	\renewcommand{\shortauthors}{Kyungho Kim et al.}
    \begin{abstract}
	Large language models (LLMs) have shown promise for training-free recommendation, but LLM-generated user interests are often too broad for fine-grained item retrieval. 
Existing methods incorporate collaborative filtering (CF) signals in a pre-LLM manner through candidate reranking or prompt augmentation, yielding limited gains.
We propose \method, a training-free recommendation framework with a post-LLM paradigm that injects CF signals into LLM-generated item representations, which are later matched with LLM-generated user interests for ranking. 
Specifically, \method
refines the directions of item embeddings using an item-item co-purchase graph and their magnitudes using item popularity.
Experiments on real-world datasets show that \method consistently outperforms existing training-free methods and achieves competitive or superior performance compared with training-based methods, without requiring any model training or task-specific fine-tuning.



    \end{abstract}
	
	\maketitle

    \section{Introduction \& Related Work}
    \label{sec:intro}
    \begin{figure}[t] 
  \centering
  \includegraphics[width=0.96\linewidth]{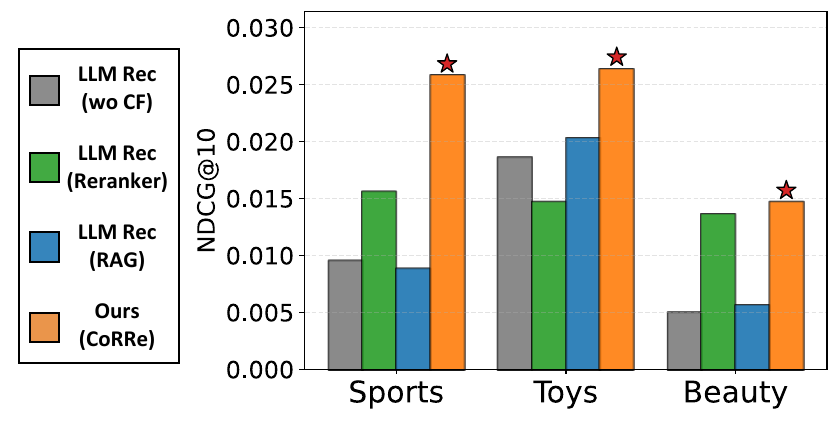} 
  \caption{
  \textit{Pre-LLM} strategies leverage collaborative signals before LLM inference, either through CF-based candidate reranking or RAG-based prompting, but show limited gains. In contrast, \method, our proposed method with a \textit{post-LLM} strategy, achieves substantially larger gains.
  }
  \label{fig:obs}
\end{figure}

Large language models (LLMs) have been widely used for recommendation~\cite{wu2024survey, kim2024large, peng2025survey, lyu2024llmrec, kim:hal-05561909} due to their strong reasoning capabilities for understanding user behavior and inferring preferences from interaction histories.
Among them, \textit{training-free} approaches, where LLMs perform zero-shot recommendation without task-specific fine-tuning, have recently gained increasing attention~\cite{lyu2024llmrec, gao2025llm4rerank, wang2024whole, kim2026itemrag}. 


However, naively leveraging LLMs for direct item recommendation remains challenging.
While LLMs can infer a user's high-level interests from interaction histories, such interests often correspond to many semantically similar items rather than a specific target item.
For example, an LLM may infer a user's interest in soccer uniforms, which still encompasses numerous products that differ only in subtle attributes such as design, color, or season.
Distinguishing among these alternatives requires fine-grained item-level discrimination beyond high-level semantic understanding.

Collaborative signals, such as co-purchase patterns and item popularity, provide valuable information for fine-grained item discrimination.
To incorporate such signals into LLMs, most existing LLM-based recommendation methods adopt a \textit{pre-LLM} strategy, where collaborative information is introduced \textit{before} the LLM makes a recommendation~\cite{wu2024survey, kim:hal-05561909}.
Specifically, collaborative filtering (CF) is first used to narrow the search space by retrieving a set of candidate items, which are subsequently reranked by the LLM~\cite{gao2025llm4rerank, kim2024large, han2025rethinking, luo2025anchor}. Alternatively, retrieval-augmented generation (RAG) approaches use CF to retrieve relevant users, items, or interaction histories and append them to the LLM prompt as additional context for recommendation~\cite{hu2026retrieval, zhang2025adaptrec, wang2025knowledge, kim2026itemrag}.

However, as shown in Figure~\ref{fig:obs}, incorporating collaborative signals in a pre-LLM manner is suboptimal, yielding only limited and inconsistent gains over LLM recommendation without any collaborative signals.
Reranking-based approaches depend heavily on CF-retrieved candidates, making the final recommendation sensitive to the quality of the initial candidate set.
RAG-based approaches incorporate collaborative signals only indirectly through retrieved context. 
As a result, collaborative signals are not fully exploited for final item selection.

To address this limitation, we propose \method, a training-free LLM recommendation framework that incorporates collaborative filtering (CF) signals in a \textit{post-LLM} manner. \method first uses an LLM to infer a user's intent toward future purchases from interaction histories and generate a user profile, which is encoded into a user embedding using an LLM embedding model~\cite{tao2024llms}.
Then, item embeddings, which are used for matching with the user embedding, are initialized by encoding item titles with the same embedding model.
To incorporate collaborative signals, \method refines the item embeddings through two complementary steps: \textit{direction refinement}, which propagates item embeddings over an item--item co-purchase graph, and \textit{magnitude refinement}, which adjusts their magnitudes according to item popularity.
Recommendations are then generated by retrieving items whose refined embeddings are most similar to the user embedding. Notably, \method remains fully training-free, requiring neither model training nor task-specific fine-tuning.

Experiments on real-world datasets demonstrate that \method consistently outperforms training-free methods and achieves performance competitive with, or superior to, training-based approaches.

Our key contributions are summarized below:
\begin{itemize}[leftmargin=*]
    \item \textbf{New paradigm.} We introduce a \textit{post-LLM} recommendation paradigm that incorporates CF signals into LLM-generated item representations for matching, rather than into LLM inputs.
    \item \textbf{New method.} We propose \method, a \textit{training-free} framework that refines LLM-generated item embeddings in two ways using co-purchase and popularity signals.
    \item \textbf{Strong performance.} \method consistently outperforms existing training-free methods and matches or surpasses training-based approaches on real-world datasets.
\end{itemize}
Supplementary materials, code, and datasets are provided at \url{https://github.com/K-Kyungho/CoRRe}.


    \section{Preliminaries}
    \label{sec:relatedwork}




\smallsection{Problem definition.} 
Let $\mathcal{U}$ and $\mathcal{I}$ denote the user and item sets, respectively. 
In the sequential recommendation setting, each user $u \in \mathcal{U}$ is represented by a chronological purchase sequence 
$\mathbf{u} \coloneqq [i^{(u)}_{1}, i^{(u)}_{2}, \cdots, i^{(u)}_{|\mathbf{u}|}]$, where $i^{(u)}_{k} \in \mathcal{I}$ denotes the $k$-th item purchased by user $u$.
Each item $i \in \mathcal{I}$ is associated with textual information (e.g., titles) $\texttt{title}_i$.
The interactions can also be represented by a binary user-item matrix $\mathbf{R}\in{0,1}^{|\mathcal{U}|\times|\mathcal{I}|}$, where $R_{ui}=1$ if user $u$ interacted with item $i$, and $0$ otherwise.
Given a user's interaction sequence $\mathbf{u}$, the goal is to rank candidate items in $\mathcal{I}\setminus\mathbf{u}$ and recommend the top-$K$ items that best match the user's preferences.

\smallsection{Embedding-based recommendation.}
Many recommendation models represent users and items as vectors in a shared embedding space, where relevance is determined by their proximity~\cite{he2020lightgcn,lee2024post}. 
Let $\mathbf{e}_u,\mathbf{e}_i\in\mathbb{R}^d$ denote the embeddings of user $u$ and item $i$, respectively.
A common choice is to measure relevance using their inner product (i.e., $\mathbf{e}_u^\top \mathbf{e}_i$).
Items with higher scores are considered more relevant and are ranked higher in the recommendation list.


    \section{Proposed Method: \textbf{\method}}
    \label{sec:method}
    \begin{figure*}[t]
    \centering
        \includegraphics[width=0.99\linewidth]{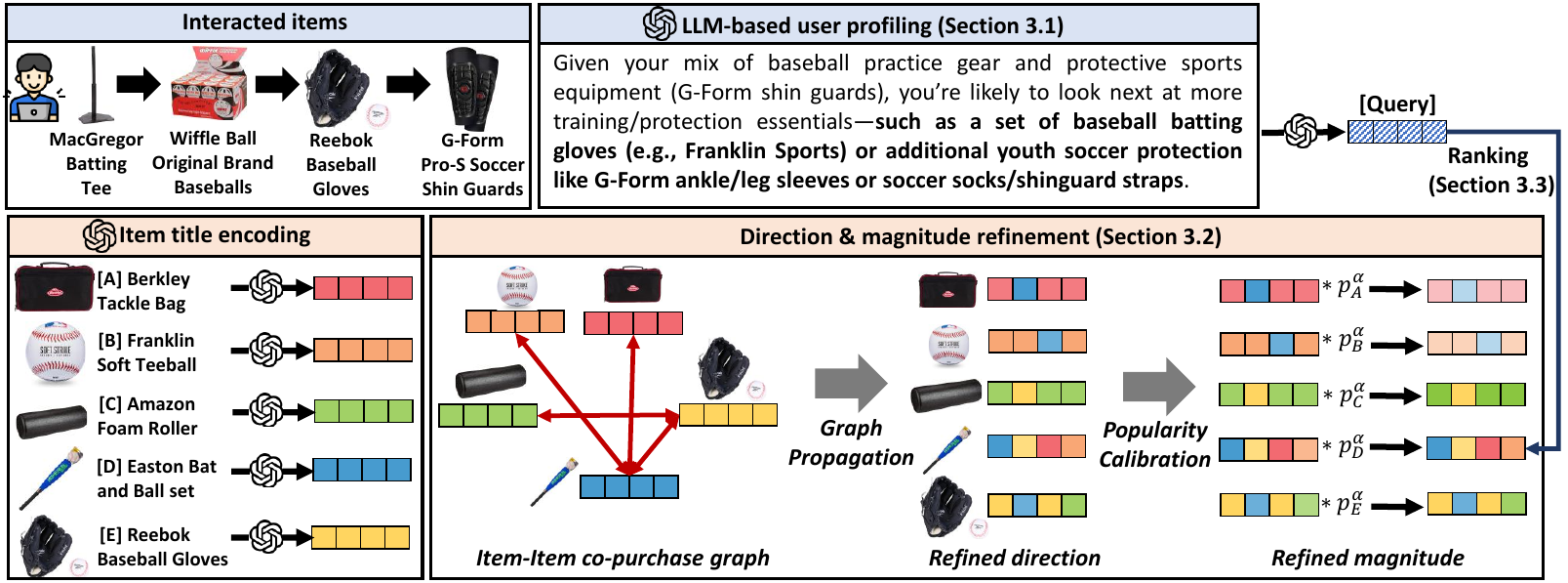}
        \caption{
        Overview of \method. \method first infers a user profile from the interaction history and uses it as a query to retrieve items from direction- and magnitude-refined item representations.
        \label{fig:model}}
\end{figure*}

We introduce a simple yet effective \textit{training-free} framework, \textbf{\method} (\textbf{\underline{Co}}llaborative \textbf{\underline{R}}etrieval \textbf{\underline{Re}}finement). 
\method consists of three components: LLM-based user profiling, item embedding refinement, and item ranking.
First, \method generates a user profile and encodes it as a query embedding.
It then refines semantic item embeddings through direction and magnitude refinement, and finally retrieves items by measuring query-item similarity.

\subsection{LLM-based User Profiling}\label{subsec:llmrec}
First, we perform LLM-based user profiling by generating a natural language profile from the user’s interaction history, capturing the user’s preferences and likely next-item interactions.
Consider a target user $u \in \mathcal{U}$ for whom we aim to generate recommendations.
For the target user, we prompt an LLM to generate a user profile $\mathcal{P}_u$ from the user’s item interaction history by inferring the user’s preferences and the types of items the user is likely to interact with next.
We provide an example prompt template in~\cite{anongithub}.
Then, we obtain the user embedding $\mathbf{e}_{u} \in \mathbb{R}^{d}$ by encoding the user profile $\mathcal{P}_{u}$ using a text encoder \texttt{Enc} (i.e., $\mathbf{e}_u=\texttt{Enc}(\mathcal{P}_u)$).
\subsection{Item Embedding Refinement}\label{subsec:retrieval}
Next, we generate item embeddings, which are later matched with the LLM-generated user embedding for ranking.
A straightforward approach is to encode each item $i\in \mathcal{I}$ using its title or description with the same text encoder as user profiles (Section~\ref{subsec:llmrec}):
\begin{equation*}
    \mathbf{e}^{\text{SEM}}_i=\texttt{Enc}(\texttt{title}_i)\in\mathbb{R}^d.
\end{equation*}
While such embeddings capture semantic information, they overlook collaborative signals, such as co-purchase patterns and item popularity, that are crucial for recommendation~\cite{he2020lightgcn,lee2024post}.

To incorporate collaborative signals into semantic item embeddings, \method applies two training-free refinement steps:
\textit{(1) direction refinement}, which injects collaborative relationships into embedding directions, and
\textit{(2) magnitude refinement}, which adjusts embedding norms according to item popularity.

\smallsection{Direction refinement.} 
We first refine the \textit{directions} of semantic item embeddings so that items with similar interaction patterns are more closely aligned in the embedding space. To this end, we construct an item-item co-purchase graph from the user-item interaction matrix by computing $\mathbf{A}=\mathbf{R}^{\top}\mathbf{R}$, where $\mathbf{A}_{ij}$ denotes the number of users who interacted with both items $i$ and $j$, indicating the \textit{strength} of their co-purchase relationship. 
Let $d_i=\sum_{j\in\mathcal{I}} \mathbf{A}_{ij}$ denote the degree of item $i$ in the co-purchase graph.
We then propagate semantic item embeddings over the co-purchase graph:
\begin{equation}\label{eq:graphprop}
    \mathbf{e}^{\text{GP}}_i
    = \sum_{j \in \mathcal{I}}
    \frac{\mathbf{A}_{ij}}{\sqrt{d_i d_j}}
    \mathbf{e}^{\text{SEM}}_j,
\end{equation}
which encourages items with strong co-purchase relationships to have more closely aligned representations.
Then, we aggregate the propagated embedding with the original semantic embedding and normalize the combined embedding:
\begin{equation*}\label{eq:direction}
    \mathbf{e}^{\text{DR}}_i= \frac{\tilde{\mathbf{e}}^{\text{DR}}_i}{\|\tilde{\mathbf{e}}^{\text{DR}}_i\|_2}
    \;\;\;\text{where}\;\;\;
    \tilde{\mathbf{e}}^{\text{DR}}_i= \lambda \mathbf{e}^{\text{SEM}}_i
    + (1-\lambda){\mathbf{e}}^{\text{GP}}_i.
\end{equation*}
Here, $\lambda$ controls the balance between the original semantic embedding and the graph-propagated embedding.


\smallsection{Magnitude refinement.}
Next, we refine the \textit{magnitude} of item embeddings.
Beyond embedding direction, magnitude also influences retrieval scores and carries important popularity information in collaborative filtering models. We therefore calibrate the norm of each direction-refined embedding according to item popularity:
\begin{equation}\label{eq:magnitude}
    \mathbf{e}^{\mathrm{DR+MR}}_i
    = p_i^{\alpha} \cdot\mathbf{e}^\text{DR}_i,
\end{equation}
where $p_i$ denotes the popularity of item $i$ (i.e., $p_i=\sum_{u\in \mathcal{U}}\mathbf{R}_{ui}$) and $\alpha$ controls the strength of popularity calibration. 
For simplicity, we denote the final refined item embedding by $\mathbf{e}_i=\mathbf{e}^{\mathrm{DR+MR}}_i$ hereafter.



\subsection{Item Ranking}
Now that we have obtained both the user embedding and the refined item embeddings, we rank items according to their similarity to the user and recommend the top-$K$ items:
\begin{equation*}\label{eq:retrieval}
\mathrm{TopK}(u)
=
\operatorname{TopK}_{i \in \mathcal{I} \setminus \mathbf{u}}
\mathrm{sim}\left(
\mathbf{e}_u,
\mathbf{e}_i
\right),
\end{equation*}
where $\mathrm{sim}(\cdot,\cdot)$ denotes a similarity function. We use the dot product in our experiments (Section~\ref{sec:experiment}).

    \section{Experimental Results}
    \label{sec:experiment}
    \begin{table*}[t]
\vspace{-1mm}
\centering
\setlength{\tabcolsep}{3pt}
\caption{\textit{(RQ1\&2) Main results \& Ablation study}. 
H@K and N@K denote Hit Ratio@K and NDCG@K, respectively. 
\gframe{\textbf{Green}} boxes and \yframe{\textbf{yellow}} boxes indicate the best and second-best results. All results are averaged over three runs.
In the ``Train-free'' column, \yes/\noo\ indicate training-free and training-based methods, respectively.
In the ``CF'' column, \yes\ denotes CF methods without LLM involvement, while \yes(\prellm) and \yes(\postllm) indicate methods that use collaborative signals before and after LLM inference, respectively. 
\method outperforms training-free baselines in 12 out of 12 cases and training-based baselines in 8 out of 12 cases.}
\label{tab:performance_tf}
{
\renewcommand{\arraystretch}{1.1}
\resizebox{1.01\linewidth}{!}{
\begin{tabular}{l c c | cccc | cccc | cccc}
\toprule
\multirow{2}{*}{Methods} & \multirow{2}{*}{\shortstack{Train\\-free}} & \multirow{2}{*}{\shortstack{CF}} & \multicolumn{4}{c|}{Sports \& Outdoors} & \multicolumn{4}{c|}{Toys \& Games} & \multicolumn{4}{c}{Beauty \& Personal Care} \\
& & & H@10 & H@20 & N@10 & N@20 & H@10 & H@20 & N@10 & N@20 & H@10 & H@20 & N@10 & N@20 \\
\midrule
\midrule

LightGCN~\cite{he2020lightgcn} & \noo & \yes
& 0.0333 & 0.0487 & 0.0206 & 0.0244 
& 0.0297 & 0.0440 & 0.0151 & 0.0187 
& 0.0257 & 0.0380 & 0.0139 & 0.0170 \\

SASRec~\cite{kang2018self} & \noo & \yes
& 0.0136 & 0.0220 & 0.0089 & 0.0108 
& 0.0108 & 0.0198 & 0.0054 & 0.0071 
& 0.0131 & 0.0225 & 0.0085 & 0.0112 \\

RecFormer~\cite{li2023text} & \noo & \yes
& 0.0263 & 0.0407 & 0.0158 & 0.0194 
& 0.0447 & 0.0627 & 0.0239 & 0.0284 
& 0.0247 & 0.0350 & 0.0124 & 0.0150 \\

RLMRec~\cite{ren2024representation} & \noo & \yes(\postllm)
& \yframe{\textbf{0.0403}} & \yframe{\textbf{0.0587}} & \yframe{\textbf{0.0229}} & \yframe{\textbf{0.0275}} 
& 0.0403 & 0.0550 & 0.0221 & 0.0257
& \gframe{\textbf{0.0363}} & \gframe{\textbf{0.0547}} & \gframe{\textbf{0.0180}} & \gframe{\textbf{0.0226}} \\

AlphaRec~\cite{sheng2025language} &\noo & \yes(\postllm)
& 0.0343 & 0.0507 & 0.0174 & 0.0215 
& \yframe{\textbf{0.0450}} & \yframe{\textbf{0.0633}} & \yframe{\textbf{0.0240}} & \yframe{\textbf{0.0286}}
& 0.0287 & 0.0417 & 0.0141 & 0.0174 \\

Rerank~\cite{singh2025gpt5} & \noo & \yes(\prellm)
& 0.0333 & 0.0487 & 0.0156 & 0.0187 
& 0.0297 & 0.0440 & 0.0147 & 0.0198 
& 0.0257 & 0.0380 & 0.0136 & 0.0175 \\

Rerank-Coral~\cite{wu2024coral} & \noo & \yes(\prellm)
& 0.0333 & 0.0487 & 0.0168 & 0.0235 
& 0.0297 & 0.0440 & 0.0144 & 0.0190 
& 0.0257 & 0.0380 & 0.0117 & 0.0165 \\

Rerank-ItemRAG~\cite{kim2026itemrag} & \noo & \yes(\prellm)
& 0.0333 & 0.0487 & 0.0158 & 0.0220 
& 0.0297 & 0.0440 & 0.0139 & 0.0188 
& 0.0257 & 0.0380 & 0.0107 & 0.0159 \\

\midrule

Popularity & \yes & \yes
& 0.0150 & 0.0200 & 0.0070 & 0.0083 
& 0.0060 & 0.0110 & 0.0025 & 0.0038 
& 0.0090 & 0.0150 & 0.0044 & 0.0058 \\

Embed-Semantic & \yes & \noo
& 0.0200 & 0.0370 & 0.0111 & 0.0153 
& 0.0360 & 0.0510 & 0.0206 & 0.0245 
& 0.0180 & 0.0220 & 0.0097 & 0.0108 \\

Embed-LLM~\cite{singh2025gpt5} & \yes & \noo
& 0.0203 & 0.0307 & 0.0095 & 0.0121
& 0.0387 & 0.0567 & 0.0186 & 0.0232 
& 0.0103 & 0.0190 & 0.0050 & 0.0071 \\

Embed-Coral~\cite{wu2024coral} & \yes & \yes(\prellm)
& 0.0183 & 0.0300 & 0.0088 & 0.0117 
& 0.0407 & 0.0597 & 0.0203 & 0.0252 
& 0.0097 & 0.0177 & 0.0056 & 0.0077 \\

Embed-ItemRAG~\cite{kim2026itemrag} & \yes & \yes(\prellm)
& 0.0163 & 0.0300 & 0.0082 & 0.0116 
& 0.0427 & 0.0607 & 0.0215 & 0.0259 
& 0.0103 & 0.0187 & 0.0058 & 0.0079 \\

\textbf{\method (Ours)} & \yes & \yes(\postllm)
& \gframe{\textbf{0.0453}} & \gframe{\textbf{0.0627}} & \gframe{\textbf{0.0258}} & \gframe{\textbf{0.0301}} 
& \gframe{\textbf{0.0507}} & \gframe{\textbf{0.0657}} & \gframe{\textbf{0.0264}} & \gframe{\textbf{0.0302}}
& \yframe{\textbf{0.0303}} & \yframe{\textbf{0.0473}} & \yframe{\textbf{0.0147}} & \yframe{\textbf{0.0189}} \\

\midrule
\midrule
\multicolumn{15}{l}{\textit{Ablation on \method}} \\
\midrule

w/o propagated emb. & \yes & \yes(\postllm)
& 0.0343 & 0.0483 & 0.0180 & 0.0214
& 0.0470 & 0.0650 & 0.0244 & 0.0289 
& 0.0167 & 0.0323 & 0.0082 & 0.0121 \\

w/o semantic emb. & \yes & \yes(\postllm)
& 0.0403 & 0.0617 & 0.0233 & 0.0289 
& 0.0233 & 0.0407 & 0.0114 & 0.0157 
& 0.0250 & 0.0387 & 0.0137 & 0.0171 \\

w/o magnitude ref. & \yes & \yes(\postllm)
& 0.0220 & 0.0400 & 0.0104 & 0.0150 
& 0.0433 & 0.0613 & 0.0217 & 0.0263
& 0.0097 & 0.0180 & 0.0057 & 0.0078 \\


\bottomrule
\end{tabular}
}
}
\end{table*}

\begin{figure*}
    \vspace{-1mm}
    \centering
    \begin{subfigure}{\linewidth}
        \centering
        \includegraphics[width=0.99\linewidth]{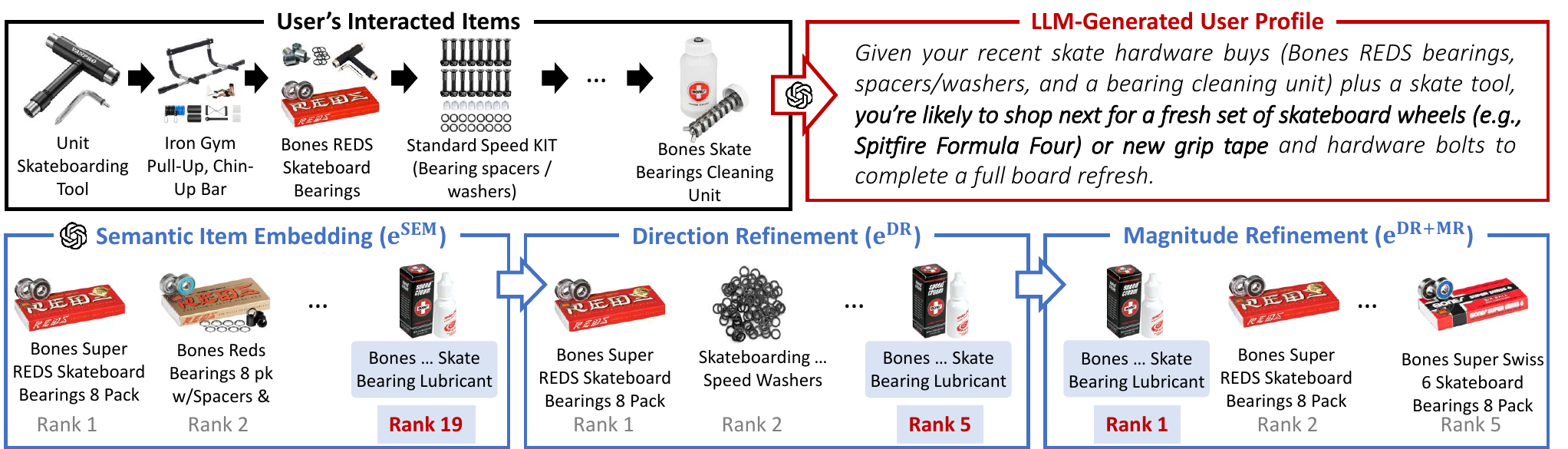}
        \caption{Sports dataset.}
        \label{fig:cs:sports}
    \end{subfigure}
    
    \begin{subfigure}{\linewidth}
        \centering
        \includegraphics[width=0.99\linewidth]{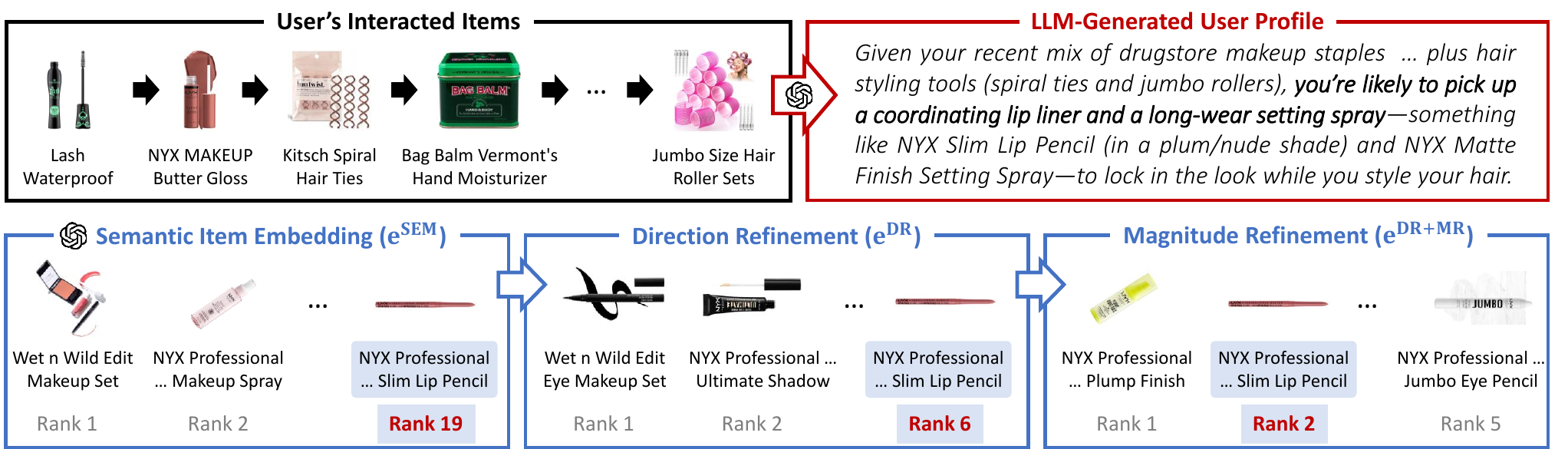}
        \caption{Beauty dataset.}
        \label{fig:cs:beauty}
    \end{subfigure}
    \caption{
    (RQ3) Case studies on two datasets. As refinement proceeds, the \bframe{ground-truth item} ranks higher among similar items.
    }
    \label{fig:cs}
\end{figure*}

In this section, we examine the following four research questions:
\begin{enumerate}[label={RQ\arabic*.}, leftmargin=*]
    \item \textbf{Accuracy:} How does \method perform compared with existing training-free and training-based recommendation methods?
    \item \textbf{Ablation study:} How does each component of \method contribute to performance?
    \item \textbf{Case study:} Does \method correctly identify the ground-truth next item beyond semantically similar alternatives?
    \item \textbf{Hyperparameter sensitivity:} How sensitive is CoRRe to the strengths of direction and magnitude refinement?
    
\end{enumerate}


\subsection{Experimental Setting}\label{subsec:protocol}

\smallsection{Datasets and evaluation.}
We evaluate \method on three domains from the latest Amazon Reviews dataset~\cite{hou2024bridging}: Sports \& Outdoors (Sports), Toys \& Games (Toys), and Beauty \& Personal Care (Beauty). 
For evaluation, we adopt the leave-one-out protocol following prior work~\cite{wu2024coral,kusano2025revisiting,kim2026itemrag},
using each user’s second-to-last purchased item for validation and the last purchased item for testing, while excluding historical items from ranking.
We randomly sample 1,000 users from each dataset and evaluate all baselines on the same sampled user set for a fair comparison. We report Hit Ratio (H@K) and NDCG (N@K) with $K \in \{10, 20\}$.

{\smallsection{Hyperparameter configurations.} \method has two key hyperparameters: $\lambda$ for direction refinement and $\alpha$ for magnitude refinement. We tune them over $\{0.2, 0.4, 0.5, 0.6, 0.8\}$ and $\{0.0, 0.05, 0.1, \\0.15, 0.2\}$, respectively, using the validation set.

\smallsection{Baselines.}
We compare \method with \textit{training-based} and \textit{training-free} methods. Training-based methods include task-specific recommenders (LightGCN~\cite{he2020lightgcn}, SASRec~\cite{kang2018self}, RecFormer~\cite{li2023text}, RLMRec~\cite{ren2024representation}, AlphaRec~\cite{sheng2025language}) and LLM reranking methods that rerank candidates retrieved by a trained recommender (Rerank, Rerank-Coral~\cite{wu2024coral}, Rerank-ItemRAG~\cite{kim2026itemrag}).\footnote{In our experiments, we rerank the top-K items retrieved by LightGCN.} 
Training-free methods include global popularity, title-based retrieval (Embed-Semantic), LLM-generated query retrieval (Embed-LLM), and their RAG-based variants (Embed-Coral~\cite{wu2024coral}, Embed-ItemRAG~\cite{kim2026itemrag}). For both reranking and embedding-based LLM baselines, we evaluate both vanilla versions using only item titles and RAG-based variants using Coral~\cite{wu2024coral} and ItemRAG~\cite{kim2026itemrag}. For all LLM-based methods, we use GPT-5.2 as the backbone LLM and text-embedding-3-large as the text encoder.

\subsection{RQ1. Accuracy}\label{subsec:standardsetting}

As shown in Table~\ref{tab:performance_tf}, \method outperforms all \textit{training-free} baselines in all settings, with up to 132.43\% improvement over the strongest baseline. The limited gains of Coral and ItemRAG suggest that simply providing additional retrieved information to the LLM before inference is insufficient for accurate recommendation.  
Furthermore, compared with \textit{training-based} methods, \method achieves the best performance on Sports and Toys and the second-best performance on Beauty, demonstrating the effectiveness of our refinement strategy without any model training. We provide additional experiments with different LLM backbones in~\cite{anongithub}.

\subsection{RQ2. Ablation Study}
We evaluate three variants of \method.
\begin{enumerate}[label=(V\arabic*), leftmargin=*]
    \item \textit{{w/o propagated embedding:}} removes the graph-propagated embedding from direction refinement, retaining only the original title-based semantic embedding.
    \item \textit{{w/o semantic embedding:}} removes the original title-based semantic embedding from direction refinement, retaining only the graph-propagated embedding.
    \item \textit{{w/o magnitude refinement:}} removes magnitude refinement.
\end{enumerate}

As shown in Table~\ref{tab:performance_tf}, (V1) and (V2) lead to a performance drop, indicating that semantic information from item titles and collaborative information from the co-purchase graph are both important. Moreover, (V3) substantially degrades performance, indicating that incorporating popularity information into the embedding magnitude is important for final retrieval.

\subsection{RQ3. Case Study}
We compare items retrieved (a) using only title embeddings, (b) after direction refinement, and (c) after direction and magnitude refinement on the Sports and Beauty datasets. As shown in Figure~\ref{fig:cs}, retrieval based solely on semantic item embeddings tends to identify items that are textually relevant to the LLM-generated user profile, but may fail to distinguish the ground-truth item from other semantically similar candidates. After direction refinement, items with stronger co-purchase relationships to the user's interacted items are promoted, improving the rank of the ground-truth item. Magnitude refinement further incorporates item popularity, helping distinguish the ground-truth item among candidates with similar semantic and collaborative relevance. 

\subsection{RQ4. Hyperparameter Sensitivity}
We analyze the sensitivity of CoRRe to two key hyperparameters: $\lambda$, which controls the balance between the original semantic embedding and the graph-propagated embedding in direction refinement, and $\alpha$, which controls the strength of magnitude refinement.

As shown in Figure~\ref{fig:hs:lambda}, the effect of $\lambda$ varies across datasets. Sports favors a smaller $\lambda$, indicating a larger benefit from collaborative information propagated through the co-purchase graph, whereas Toys benefits from a larger $\lambda$, placing more emphasis on the original semantic representation. Similarly, Figure~\ref{fig:hs:alpha} shows dataset-dependent behavior for $\alpha$. Sports benefits from a moderate degree of popularity calibration, while Toys and Beauty favor smaller values of $\alpha$, suggesting that overly emphasizing popularity can degrade retrieval performance.

    \section{Conclusion}
    \label{sec:conclusion}
    In this work, we propose \method, a training-free recommendation framework that separates user intent inference from item-level retrieval. 
\method incorporates collaborative signals into item representations in a post-LLM manner through direction and magnitude refinement.
Experiments demonstrate that
\method consistently outperforms existing training-free recommenders and achieves competitive or superior performance compared with training-based recommenders.
For reproducibility, we provide our code, datasets, and supplementary materials at \url{https://github.com/K-Kyungho/CoRRe}.

\begin{figure}[t]
    \centering
    \begin{subfigure}{\linewidth}
        \centering
        \includegraphics[width=0.99\linewidth]{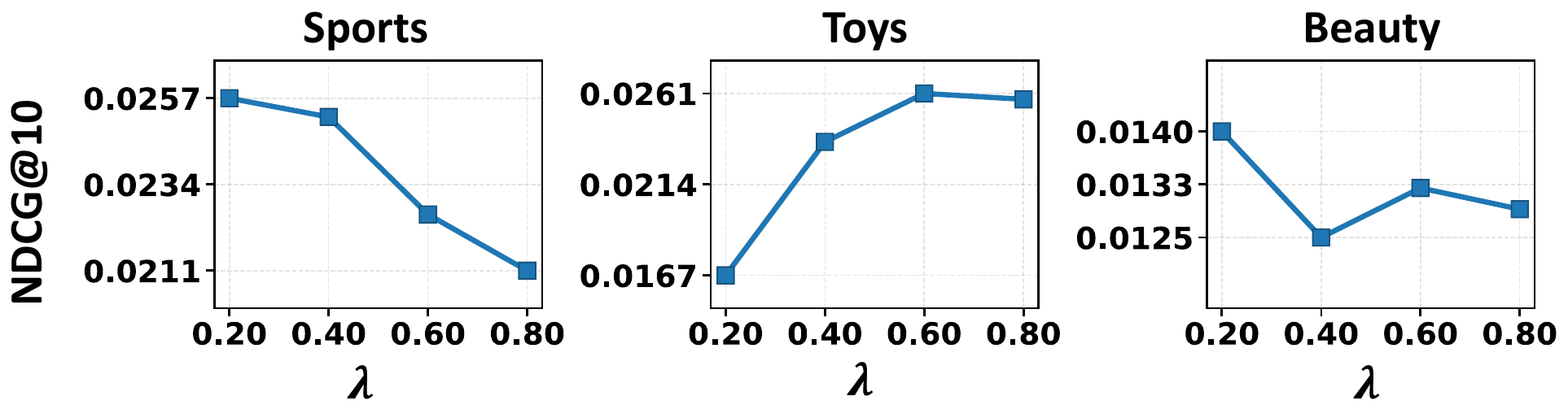}
        \caption{Hyperparameter sensitivity to $\lambda$.}
        \label{fig:hs:lambda}
    \end{subfigure}
    
    \begin{subfigure}{\linewidth}
        \centering
        \includegraphics[width=0.99\linewidth]{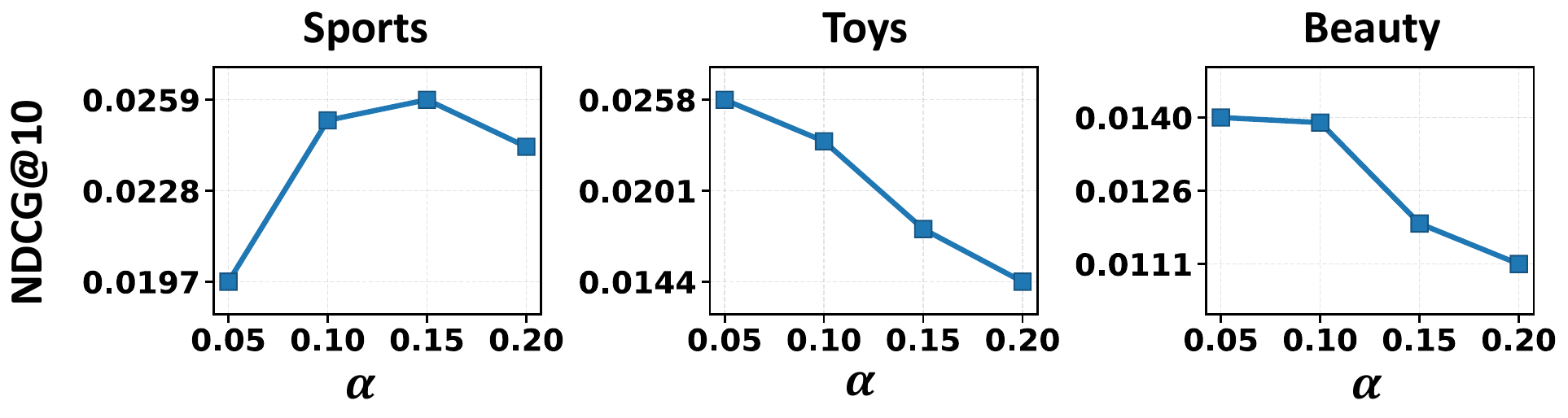}
        \caption{Hyperparameter sensitivity to $\alpha$.}
        \label{fig:hs:alpha}
    \end{subfigure}
    \caption{
    (RQ4) Hyperparameter sensitivity of \method to (a) $\lambda$, which controls the balance between semantic and graph-propagated item embeddings, and (b) $\alpha$, which controls the strength of popularity-based magnitude calibration.}
    \label{fig:hs}
\end{figure}


    \section*{Acknowledgements.}
    This work was partly supported by the National Research Foundation of Korea (NRF) grant funded by the Korea government (MSIT) (No. RS-2024-00406985, 30\%).
    This work was partly supported by Institute of Information \& Communications Technology Planning \& Evaluation (IITP) grant funded by the Korea government (MSIT) (No. RS-2024-00438638, EntireDB2AI: Foundations and Software for Comprehensive Deep Representation Learning and Prediction on Entire Relational Databases, 30\%) (No. RS-2024-00457882, National AI Research Lab Project, 30\%)
    (No. RS-2019-II190075, Artificial Intelligence Graduate School Program (KAIST), 10\%).

    \section*{GenAI Usage Disclosure}
    We used LLM-based tools solely to assist with minor writing edits and code debugging. All generated outputs were carefully reviewed and revised by the authors to ensure accuracy and integrity.

    \bibliographystyle{ACM-Reference-Format}
    \balance
	\bibliography{000Ref}
    
    







\end{document}